\documentclass[aps,prl,twocolumn,showpacs,preprintnumbers,amsmath,amssymb]{revtex4}
\usepackage{graphicx}% Include figure files
\usepackage{dcolumn}% Align table columns on decimal point
\usepackage{mathrsfs}
\usepackage{bm}% bold math
\usepackage{amsfonts}
\usepackage{amssymb}
\usepackage{amsmath}
\usepackage{color}
\begin{document}

\preprint{}

\title{Electromagnetically Induced Transparency and Quantum Heat Engines}

\author{S.E. Harris}
\email{seharris@stanford.edu}
\affiliation{Departments of Electrical Engineering and Applied Physics, Ginzton Laboratory, Stanford University, Stanford, California, 94305, USA}
\date{\today}
\begin{abstract}
We describe how  electromagnetically induced transparency may be used to construct a non-traditional near-ideal quantum heat engine as constrained by the Second Law. 
The engine is pumped by a thermal reservoir that may be either hotter or colder than that of an exhaust reservoir, and also by a monochromatic laser. As output, it produces a bright narrow emission at line center of an otherwise absorbing transition. 
\end{abstract}
                              %display desire
\maketitle

\begin{figure}[tbp]
\begin{center}
\includegraphics[width=9.0cm]{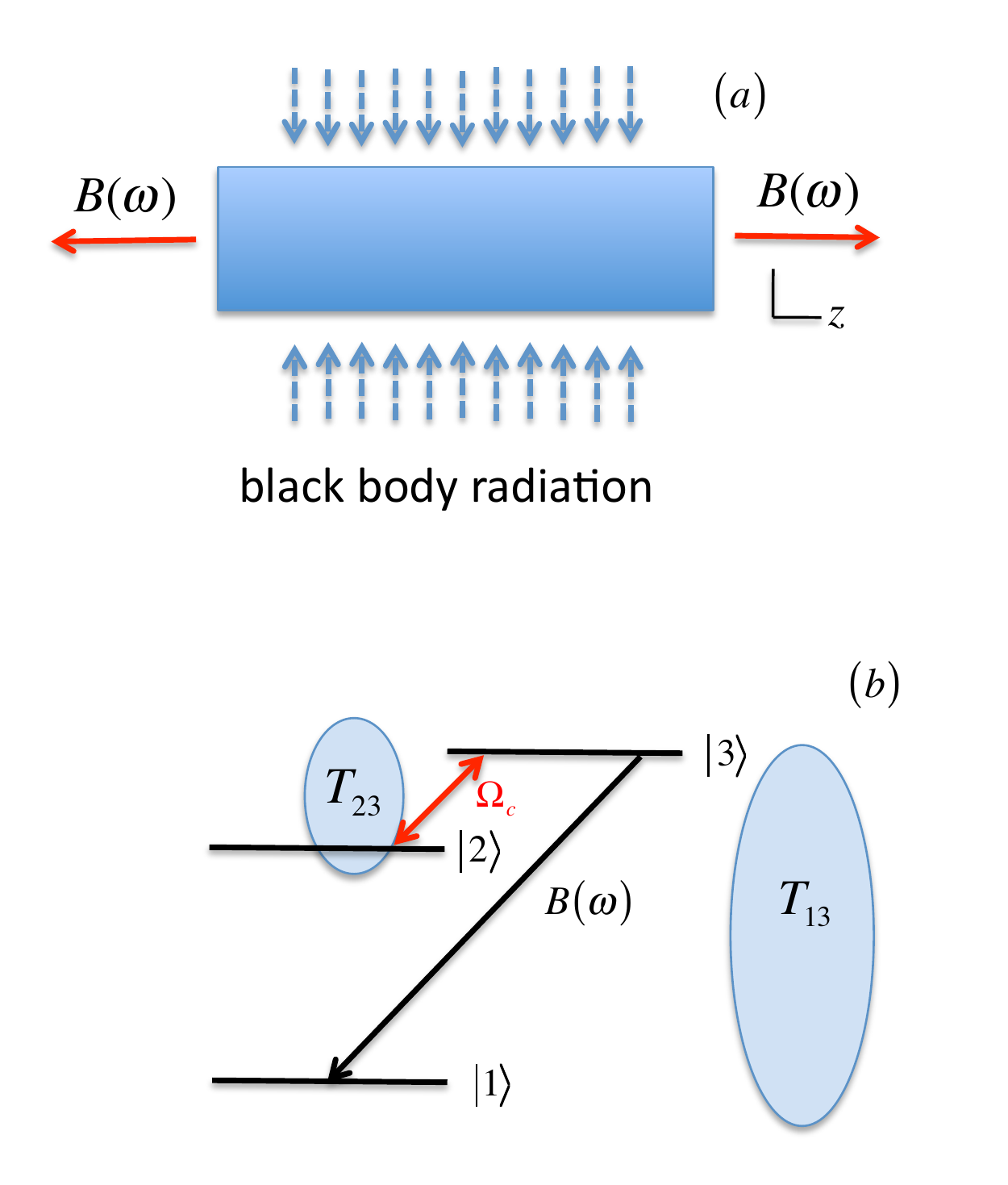}
\end{center}
\caption{ An ensemble of atoms is  pumped by black body radiation. The pumping temperature is  $T_{13}$ on the  $\left | 1 \right > \rightarrow \left|3\right >$ transition and  $T_{23}$ on the $\left | 2 \right > \rightarrow \left |3\right >$  transition. A monochromatic ``coupling'' laser with Rabi frequency $\Omega_c$ is tuned to resonance of the 
$\left | 2 \right > \rightarrow \left|3\right >$ transition.  As a result of EIT there is a spectrally narrow and bright emission in the $z$ direction. This emission will generally have a higher temperature than that of either of the pertinent reservoirs (color online).}
\label{fig:schematic}
\end{figure}

There is substantial on-going interest in quantum heat engines and refrigerators both for possible application and as part of the study of quantum thermodynamics \cite{Kosloff1}. This work began decades ago when Scovil et al. described the equivalence of a three level maser and a Carnot engine \cite{Scovil,Geusic}. In recent years the scope of this work was significantly expanded when Scully and colleagues recognized that by using a microwave or laser source to introduce coherence and overcome ``Golden rule" physics that substantial improvements in performance become possible \cite{Scully1, Scully2, Scully3}. Work in the context of tricycle heat engines has followed and it has been recognized that the second law allows, as described in the following text, unique capabilities \cite{Kosloff2, Rozek, Kurizki, Ivanov} .  

Following in this direction, this Letter describes  the use of electromagnetically induced transparency (EIT) to construct a near-ideal but non-traditional quantum heat engine. EIT is used to establish absorptive and emissive profiles that have near zero absorption at the wavelength of maximum emission \cite{FleischhauerReview, Kocharovskaya, HarrisNonlinear, HarrisPhysicsToday, ScullyBook}. When this system is driven by black body radiation, Fig. 1(a),  because of the transparency, the radiation $B(\omega)$ in the $\pm$ z-direction is much stronger than that predicted by Kirchhoff's law. This radiation has non-zero entropy, may drive a piston, and is in the category of low grade work.

Figure 1(b) shows the atomic system. The $\left | 1 \right > \rightarrow \left|2\right >$ transition is metastable with transition frequency $\omega_{12}$. A monochromatic ``coupling"  laser with Rabi frequency $\Omega_c$ is applied at line center of  the $\omega_{23}$  transition. Black body radiation at a temperature $T_{13}$ interacts with the $\left | 1 \right > \rightarrow \left|3\right >$ transition and blackbody radiation at a temperature $T_{23}$ interacts with the $\left | 2 \right > \rightarrow \left |3\right >$ transition. These temperatures may be different or the same, and either may be higher than the other. We assume filters on the pumping radiation that are not shown in Fig.(1). Phase matching is not involved and the directions of all of the fields are arbitrary. 

In terms of the flow of input and output photons, the overall process is as follows:  For each photon that is absorbed from the $T_{13} $ reservoir a photon is generated on the $T_{23}$ reservoir, thereby increasing the population of state  $\left | 2 \right >$ by one unit.  A photon is then absorbed from the coherent coupling laser (red) on the 
$\left | 2 \right > \rightarrow \left|3\right >$ transition and a photon is generated on the $\left | 3 \right > \rightarrow \left|1\right >$ transition. Of note, as in nonlinear optical processes governed by the Manley-Rowe relations \cite{Manley-Rowe}, power trades in units of photons; i.e. only a fraction $\omega_{23}/  \omega_{13}$ of the power generated at $\omega_{13}$ comes from the coupling laser.

The atomic system of Fig.1(b) has been previously studied by Imamoglu, Field, and Harris \cite{ImamogluClosedPRL} as a prototype closed system for lasers that do not require a population inversion.  Of importance, because the system is closed, the driving black body excitation rates and the dephasing rates are determined by the lifetime decay rates $\Gamma_{31} $ and $\Gamma_{32} $ and the ambient temperatures. The rates $R_{ij}=R_{ji}$ are related to the thermal occupation numbers $\bar{n}_{13}$ and $\bar{n}_{23}$ by
\begin{eqnarray}
&R_{23}=\Gamma_{32} \bar{n}_{23}=\Gamma_{32} \{ \exp[\hbar \omega_{23}/k_{b}T_{23}]-1\}^{-1}\nonumber\\
&R_{13}=\Gamma_{31} \bar{n}_{13}=\Gamma_{31} \{ \exp[\hbar \omega_{13}/k_{b}T_{13}]-1\}^{-1}
\end{eqnarray}
The dephasing rates of each of the transitions are
\begin{eqnarray}
&&\gamma_{21}=R_{23}+R_{13} \nonumber\\
&&\gamma_{31}=\Gamma_{31}+\Gamma_{32}+R_{23}+2 R_{13} \nonumber\\
&&\gamma_{32}=\Gamma_{31}+\Gamma_{32}+ R_{13}+2 R_{23}
\end{eqnarray}
Therefore, once an atomic system is chosen, the only free parameters are the Rabi frequency of the coupling laser and the ambient pumping temperatures $T_{13}$ and $T_{23}$. 
Assuming that there is no reflection or scattering into other modes, we calculate the spectral brightness $B(\omega, z)$ for a single transverse mode as a function of distance $z$.  
$B(\omega, z)$ is dimensionless so that the number of photons per second that are generated in the $z$ direction is $1/(2 \pi) \int B(\omega, z) d\omega$. 

With $N$ as the atom density, $\rho_{ii}$, as the diagonal density matrix elements, and absorption and emission cross sections $\sigma_{abs}$ and $\sigma_{em}$, the equation\cite{Mahalas} for $B(\omega, z)$ is
\begin{eqnarray}
&&\frac{dB(\omega, z)}{dz}+N \left[\sigma_{abs} \rho_{11}-\sigma_{em} (\rho_{22}+\rho_{33}) \right] B(\omega, z)= \nonumber \\
&&\sigma_{em} (\rho_{22}+\rho_{33}).  
\end{eqnarray}

The spectral brightness is zero at $z=0$ and at $z$ sufficiently large that all spectral components of interest are absorbed reaches a maximum value of $B_{black}(\omega)$. 
With $\Lambda$ defined as the ratio of  atoms in the upper manifold to those in ground; i.e. $\Lambda=(\rho_{22}+\rho_{33})/  \rho_{11}$ the limiting brightness at each spectral component is 
\begin{equation}
B_{black}(\omega)=\frac{\Lambda\   \sigma_{em}}{\sigma_{abs}-\Lambda\  \sigma_{em}}
\end{equation}

We assume that field on the $\left | 1 \right > \rightarrow \left|3\right >$  transition is sufficiently weak that the populations are determined by the driving rates $R_{ij}$ and the strong coupling field $\Omega_{c}$.  Solving for $\rho_{ii}$, $\Lambda$ is 
\begin{equation}
\Lambda=\frac{R_{13} \left(2 \Omega _c^2+\gamma _{32} \left(\Gamma _{32}+2
   R_{23}\right)\right)}{\left(\Gamma _{31}+R_{13}\right) \left(\Omega _c^2+\gamma _{32}
   R_{23}\right)}
\end{equation}

Following Imamoglu et al., the expressions for the absorptive and emissive cross sections are obtained by solving the density matrix equations with the assumption that the Rabi frequency of the coupling laser is large as compared to all other Rabi frequencies in the system \cite{ImamogluClosedPRL, Imamoglu}. With $\omega$ as the frequency of the generated spontaneous radiation and  $\Delta\omega=\omega_{13}-\omega$, these cross sections are 
 \begin{widetext}
\begin{eqnarray}
&&\frac{\sigma_{abs}}{\sigma_0}=\frac{\gamma _{31} \left(\gamma _{21} \Omega _c^2+\gamma _{31} \left(\gamma _{21}^2+4
   \Delta \omega ^2\right)\right)}{4 \Delta \omega ^2 \left(-2 \Omega _c^2+\gamma
   _{21}^2+\gamma _{31}^2\right)+\left(\Omega _c^2+\gamma _{21} \gamma
   _{31}\right){}^2+16 \Delta \omega ^4}\nonumber\\
   and
&&\frac{\sigma_{em}}{\sigma_0}=\frac{\gamma _{31} \Gamma _{32} \Omega _c^2 \left(\Omega _c^2+\gamma _{21} \gamma
   _{31}-4 \Delta \omega ^2\right)+\gamma _{31} \left(\gamma _{21} \left(\Omega
   _c^2+\gamma _{21} \gamma _{31}\right)+4 \gamma _{31} \Delta \omega ^2\right)
   \left(\Omega _c^2+\gamma _{32} R_{23}\right)}{\left(4 \Delta \omega ^2 \left(-2
   \Omega _c^2+\gamma _{21}^2+\gamma _{31}^2\right)+\left(\Omega _c^2+\gamma _{21}
   \gamma _{31}\right){}^2+16 \Delta \omega ^4\right) \left(\gamma _{32} \Gamma
   _{32}+2 \Omega _c^2+2 \gamma _{32} R_{23}\right)}
\end{eqnarray}
\end{widetext}
Both cross sections are normalized to $\sigma_{0}=(2 \omega_{13} |\mu_{13} |^2)/(\epsilon_{0} c  \hbar \gamma_{13})$ 
where $\mu_{13}$ is the transition matrix element. For a lifetime  broadened transition  in an isotropic medium $\sigma_{0}=\lambda^2/ 2 \pi$.
The brightness is given by Eq.(4), with the population ratio and the absorptive and emissive cross sections given by Eqs (5-6). 
\begin{figure}[tbp]
\begin{center}
\includegraphics[width=9.5cm]{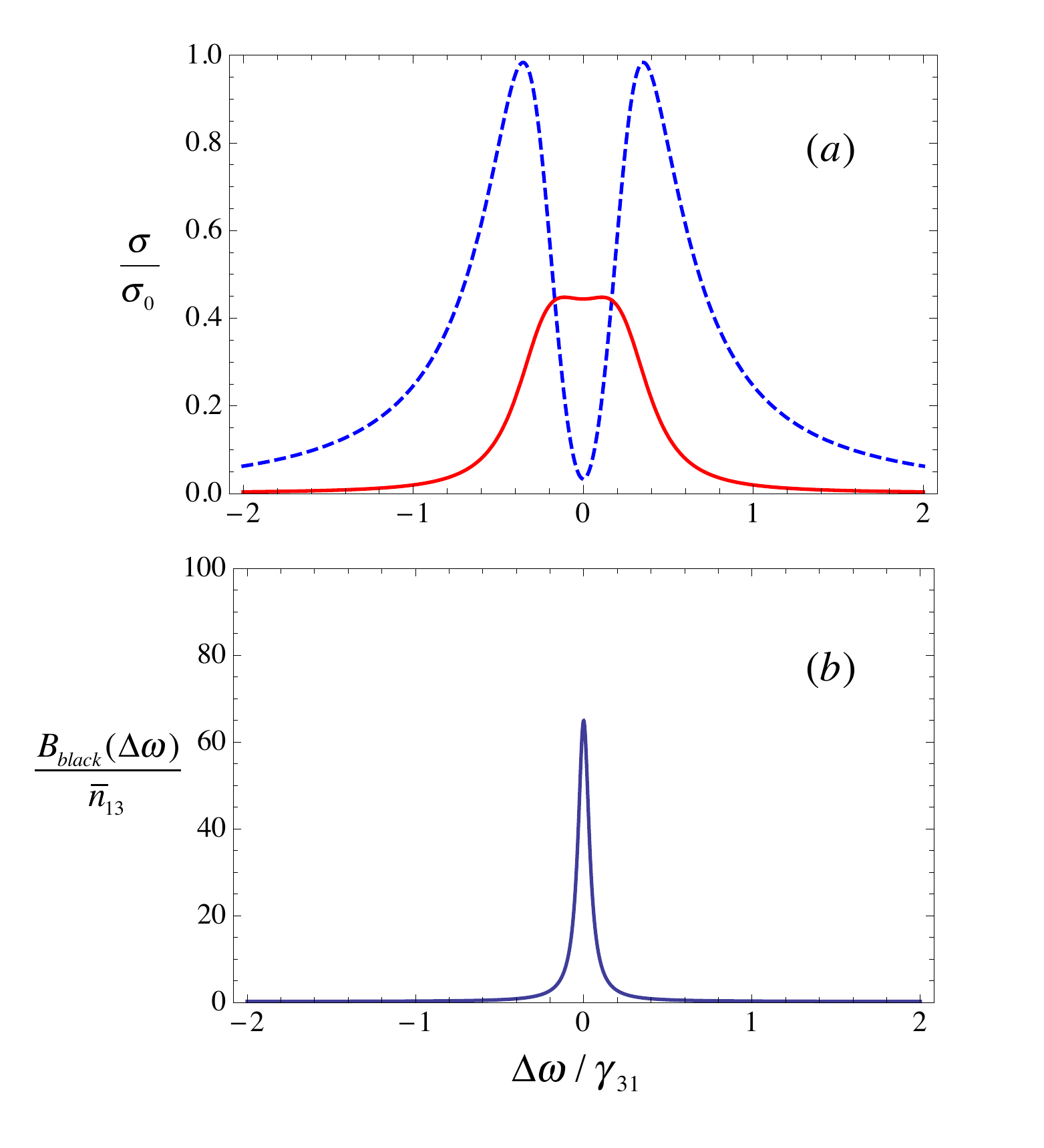}
\end{center}
\caption{ (a) Absorptive (dashed line) and emissive (solid line) cross sections as a function of the detuning from resonance $\Delta \omega$ . (b) Spectral brightness $B_{black}(\Delta\omega)$ normalized to the modal number $\bar{n}_{13}$. At peak the normalized brightness is $64.9$ times larger than that of Kirchhof's law.  Parameters  are  $\Gamma_{31}=10^{7}$, $\Gamma_{32}=6 \times10^{7}, \Omega_{c}=5 \times10^{7} , \omega_{13}= 4\times 10^{15},  \omega_{12}=10^{15}$ and $ T_{0}=5778^\circ  K $. The population ratio $\Lambda$ is 0.019.}
\end{figure}

Figure 2 (a) shows the cross sections for absorption and for emission, both as a function of the detuning from resonance $\Delta\omega$ . In Fig. 2 (b) we assume a common pumping temperature $T_{13}=T_{23}=T_{0}$ and plot $B_{black}(\Delta \omega)$.  As anticipated there is a narrow and strong emission at line center. [Equivalently, one may solve Eq. (3) with an optical depth of at least $N \rho_{11}\sigma_{abs} L=10$ at all spectral components of interest to obtain nominally the same result as obtained with Eq. (4).]  A point of caution: There are  tails in the spectral brightness whose magnitude is determined by the optical depth that are not visible on the scale of Fig.~ 2(b). For example at an optical depth at line center of $10$ in the $z$~direction, the ratio of the peak brightness $B_{black}(\Delta \omega=0)$ to the brightness $B_{black}(\Delta \omega=\pm 10\ \gamma_{31})$ is $554$.

 Combining previous equations the brightness at line center $\Delta \omega=0$ is
\begin{eqnarray}
B_{black}(0) =-\frac{\bar{n}_{13} \left(\gamma _{21} \gamma _{32} \Gamma _{32}
   \bar{n}_{23}+\left(\gamma _{21}+\Gamma _{32}\right) \Omega _c^2\right)}{\Gamma
   _{32} \bar{n}_{13} \Omega _c^2-\gamma _{21} \left(\gamma _{32} \Gamma _{32}
   \bar{n}_{23}+\Omega _c^2\right)}
\end{eqnarray}
For the atomic system of  Fig. (1) there are two Feynman paths that are, independently, in detailed balance. The first is the single photon path $\left |1 \right > \leftrightarrow \left|3\right >$. When there is no coupling laser, this is the only path and the spectral brightness $B_{black}(0)=\bar{n}_{13}$ and the equivalent temperature is $T_0$.  The second path is the two photon path $ \left | 1 \right >\rightarrow \left|3\right >\rightarrow \left|2\right >$ where the transitions into and out of state $\left | 3 \right >$ are virtual. At large Rabi frequency 
$\Omega_c$  this latter path is isolated and the brightness of Eq. (7) approaches 
\begin{equation}
\lim \left[B_{black}(0), \Omega_{c}\rightarrow \infty \right] =\frac{\bar{n}_{13} \left(\Gamma _{31} \bar{n}_{13}+\Gamma _{32}
   \left(\bar{n}_{23}+1\right)\right)}{\Gamma _{31} \bar{n}_{13}+\Gamma _{32}
   \left(\bar{n}_{23}-\bar{n}_{13}\right)}
\end{equation}
Figure (3) illustrates this behavior. Part (a) shows the line center value of the normalized brightness as a function of the Rabi frequency of the coupling laser.  Part (b) shows this same quantity, but now plotted as the normalized temperature $T/T_{0}$ that is in correspondence with the spectral  brightness; i.e.  $T=(\hbar \omega_{13}/k )/\left[\ln(1/B+1)\right]$. 
\begin{figure}[tbp]
\begin{center}
\includegraphics[width=9.5cm]{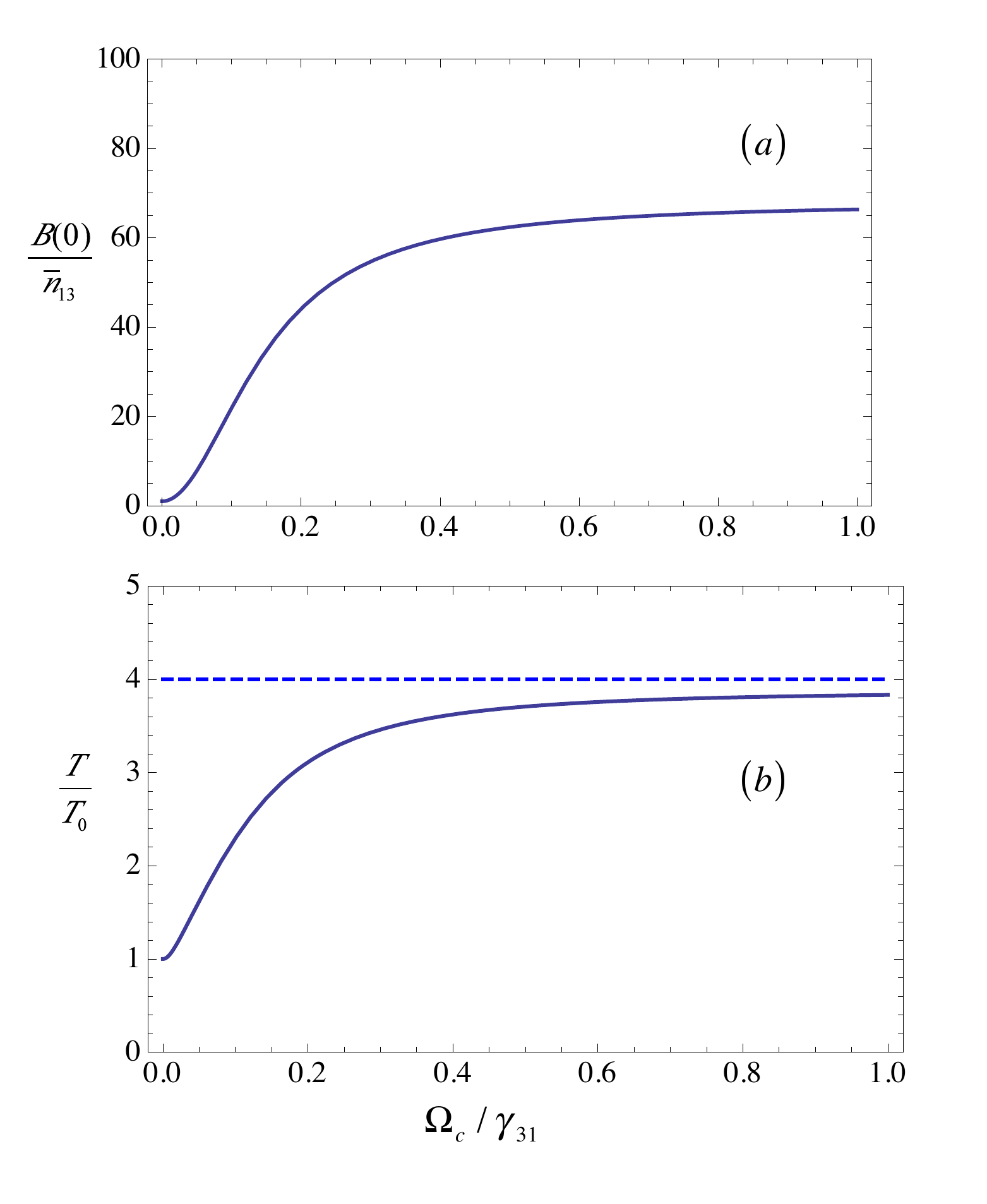}
\end{center}
\caption{(a) $B_{black}(\Delta \omega=0)$ versus $\Omega/\gamma31$ normalized to  $\bar{n}_{13}$ and (b) the same quantity in units of normalized temperature.  The parameters are the same as those of Fig. 2. The dashed line is the temperature $T_{B} $ predicted by the second law.}
\end{figure}
As the strength of the coupling laser is increased the spectral  brightness and equivalent temperature rise smoothly from their initial values $\bar{n}_{13}$ and $T_{0} $ to the limiting value of  Eq.(8).  The dashed line is the temperature as predicted by Eq. (10), and as developed in the next paragraph is greater than $T_{0}$ in the same ratio as $\omega_{13}/\omega_{12}$.

We now compare the EIT based heat pump to an ideal pump. From Fig. 1 (b). For each  photon taken from the  $\left | 1 \right > \rightarrow \left |3\right >$ reservoir, a photon is generated in the $\left | 2 \right > \rightarrow \left |3\right >$ reservoir. The coupling laser loses a photon to produce a photon along the z axis with a temperature $T_{B}$. Noting that the entropy of a monochromatic laser is unchanged by losing or gaining a photon \cite{Mungan, Ruan}, the requirement for increasing total entropy is \cite{Geusic, Field}
\begin{eqnarray}
&&\Delta S=-\frac{\hbar \omega_{13}}{T_{13}}+\frac{\hbar \omega_{23}}{T_{23}}+\frac{\hbar \omega_{13}}{T_{B} }\geq 0 \nonumber \\
&&T_{B}\leq \frac{T_{13} T_{23}\  \omega _{13}}{T_{23}\  \omega _{13}-T_{13} \ \omega _{23}}
\end{eqnarray}
In Fig. (3b), we have $T_{13}=T_{23}=T_{0}$, and therefore $T_{B}=(\omega_{13}/(\omega_{13}-\omega_{23}) T_{0}=(\omega_{13}/(\omega_{12}) T_{0}$. The dashed line is this value. We have thereby assumed that $B_{black}(\omega)$ is in the category of low grade work and has the same entropy as a filtered thermal beam \cite{Ruan}. 

With one more approximation, $\Gamma_{31}<<\Gamma_{32}$, Eq.(8) becomes
\begin{eqnarray}
&&B_{max}=\left( \frac{\bar{n}_{23}+1}{\bar{n}_{23}-\bar{n}_{13}}\right) \bar{n}_{13}\nonumber\\ 
&&T_{max}=\frac{T_{13} T_{23}\  \omega _{13}}{T_{23}\  \omega _{13}-T_{13} \ \omega _{23}}
\end{eqnarray}
where in the last step we used  $T_{max}=(\hbar \omega_{13}/k )/\left[\ln(1/B_{max}+1)\right]$. Therefore, in the appropriate limits, the peak generated brightness is equal to that which is allowed by the second law. Except when close to the singularity in the denominator of Eq.(9), the approximation $\Gamma_{31}<<\Gamma_{32}$ is not stringent. For example, in Fig. 3(b), for reasonably general parameters, the maximum value of $B(0)$ is approached. The condition that  the denominator in Eq (9) or Eq (10) be positive sets limits on the value of the temperature of the reservoirs where the previous formuli apply. For example, if $T_{13}$ is chosen, then $T_{23}$ must lie in the range $(\omega_{23}/\omega_{13})  T_{13}<T_{23}<\infty$.

Of importance, when $\Gamma_{32}>\Gamma_{31}$, the singularity $T_{23}\  \omega _{13}= T_{13}\  \omega _{23}$ or equivalently $\bar{n}_{23}=\bar{n}_{13}$ denotes the threshold for lasing without population inversion \cite{ImamogluClosedPRL}.  As in a normal unsaturated laser, at or above this threshold, the temperature $T_{B}$ [Eq.(9)] will increase indefinitely and we may compare the efficiency of the EIT based engine to the prototype engine of Scovil and  Geusic et. al. \cite{Scovil, Geusic}.  

For normalization we first consider the energy level diagram of Fig. 1(b), but \textit {with the coupling laser turned off}. We assume that the $\left | 1 \right > \rightarrow \left|2\right >$ transition is metastable with a small, but non-zero, matrix element. Defining the  efficiency of a laser on this transition as the ratio of the per photon output energy to the input energy, the efficiency is 
\begin{eqnarray}
\eta_{Carnot}&&=\frac{\omega_{12}}{\omega_{13}}=\frac{\omega_{13}-\omega_{23}}{\omega_{13}}\nonumber\\
&&=1-\frac{T_{23}}{T_{13}}=1-\frac{T_{C}}{T_{H}}
\end{eqnarray}
Here, the second equality follows from the population inversion condition $\omega_{23}/T_{23}=\omega_{13}/T_{13}$ and taking $T_{23}$ as the colder of the two reservoirs.

Proceeding as above the efficiency of the EIT based engine when at  threshold  is
 \begin{eqnarray}
\eta=\frac{\omega_{13}}{\omega_{13}+\omega_{23}}\nonumber\\
=\frac{T_{13}}{T_{13}+T_{23}}
\end{eqnarray}
The first equality in Eq. (12) follows by taking the input energy as the sum of the energy of a photon from the  $\omega_{13}$ reservoir and a photon from the coupling laser, and the output energy as that of the generated photon at $B(\Delta\omega=0)$. The second equality in Eq. (12) follows from the singularity in the denominator of Eq.(10); i.e. operation at  reservoir temperatures that are at threshold for lasing without population inversion \cite{Fry}. 

There are substantial differences between the efficiency of the traditional Scovil-Geusic Carnot engine and the EIT based engine. Perhaps the most important of these is that, for the EIT engine, it is not required that $T_{13}>T_{23}$, and Eq.(12) holds for both cases.  When $T_{13}$ is greater than $T_{23}$ then, at threshold, the ratio of the efficiency of the EIT based engine generating radiation on the $\left | 1 \right > \rightarrow \left|3\right >$ transition to the Carnot engine operating on the  $\left | 1 \right > \rightarrow \left|2\right >$ transition is
 \begin{eqnarray}
\frac{\eta}{\eta_{Carnot}}=\frac{T_{13}^2}{T_{13}^2-T_{23}^2}
\end{eqnarray}
This ratio is always greater than unity. 

In summary, quantum heat pumps and electromagnetically induced transparency are well known terms in quantum optics. This work has established a strong connection. 
We have also shown how, using the Second Law, one may easily obtain a result, i.e. Eq.(9), that using Maxwell's and Schr\u{o}dinger's equations takes several pages of calculation, i.e. Eq.(10). 
  
The author acknowledges many contributions, made long ago, of Atac Imamo\u{g}u and John Field to his  work in this field. A discussion of the possibility of ultra-bright spontaneous emission using EIT and of the role of entropy is given in J. E. Fields Ph.D. dissertation \cite{Field}. I thank Shanhui Fan for suggesting that I consider brightness enhancement in the context of heat engine physics, and thank Olga Kocharovskaya, David Miller, and Marlan Scully for helpful discussions. I also thank M. Scully for showing that the change in entropy, as in Eq. (9), is exact for thermal beams.

%\bibliography{monolithic}

\end{document}